\documentclass{jpconf}
\usepackage{epsfig}
\usepackage{amsmath}
\begin{document}
\newcount\eLiNe\eLiNe=\inputlineno\advance\eLiNe by -1
\title{Transverse Momentum Distributions in p-Pb collisions and Tsallis Thermodynamics.}
%
% subtitle is optionnal
%
%%%\subtitle{Do you have a subtitle?\\ If so, write it here}
\author{M. D. Azmi and J. Cleymans}
\address{UCT-CERN Research Centre and Department  of  Physics,\\ University of Cape Town, Rondebosch 7701, South Africa}
\date{\today}
\begin{abstract}
The transverse momentum distributions of charged particles in p-Pb collisions
at $\sqrt{s_{NN}}$ = 5.02 TeV  measured  by the ALICE collaboration are fitted using Tsallis statistics. 
The use of a thermodynamically consistent
form of this distribution leads to an excellent description of the transverse momentum distributions for all
rapidity intervals.
The values of the Tsallis parameter $q$, the temperature $T$ and the radius $R$ of the system 
do not change within the measured  pseudorapidity interval.
\end{abstract}
%
%
%
%\pacs{} \keywords{p-Pb, Tsallis, transverse momenta}
%
%\maketitle
%
\section{\label{sec:Introduction}Introduction}
It is by now standard to parameterize transverse momentum distributions
with functions having a power law behavior at high momenta. This has been done 
by the  STAR~\cite{Abelev:2006cs} and PHENIX~\cite{Adare:2010fe,Adare:2011vy} collaborations at RHIC and by the  
ALICE~\cite{Aamodt:2011zza,Aamodt:2011zj}, ATLAS~\cite{Aad:2010ac} and CMS~\cite{Khachatryan:2010us,Chatrchyan:2012qb} collaborations at 
the LHC using a form of the Tsallis~\cite{Tsallis:1987eu} distribution.
This distribution was  proposed as a generalization of the Boltzmann-Gibbs distribution  and introduces a new parameter $q$. In the limit $q = 1$
it coincides with the latter. It has been shown repeatedly that the Tsallis distribution gives an excellent description of $p-p$
collisions~\cite{Cleymans:2011in,Sena:2012ds,Wong:2013sca} resulting in a value for the $q$ parameter which is around 1.1 - 1.2.
In this paper we extend this analysis to transverse momentum spectra obtained in $p-Pb$ collisions at $\sqrt{s_{NN}}$ = 5.02 TeV by the 
ALICE collaboration~\cite{ALICE:2012mj}.  We conclude that the Tsallis distribution gives an excellent fit to $p-Pb$ collisions 
 with values for $q$ and $R$ which are consistent with those obtained in  $p-p$ collisions but with a temperature 
which is clearly higher, around 112 MeV.

\section{Tsallis Distribution}
\indent In the framework of Tsallis 
statistics~\cite{Cleymans:2011in,Tsallis:1987eu,Biro:2008hz,Conroy:2010wt,Cleymans:2012ya}
the entropy $S$, the particle number, $N$,  the energy density $\epsilon$ and the pressure $P$ 
are given by 
integrals over the  Tsallis distribution: 
\begin{equation}
f = \left[ 1 + (q-1) \frac{E-\mu}{T}\right]^{-\frac{1}{q-1}} .\label{tsallis} 
\end{equation}
\indent It can be shown (see e.g.~\cite{Cleymans:2012ya}) that the relevant thermodynamic quantities are given by:
\begin{eqnarray}
	S&=&-gV\int\frac{d^3p}{(2\pi)^3}
\left[ f^{q}\ln_{q} f -f \right],
\label{entropy} \\
N &=& gV\int\frac{d^3p}{(2\pi)^3} f^q ,\label{number} \\
\epsilon &=& g\int\frac{d^3p}{(2\pi)^3}E f^q ,\label{epsilon}\\
P &=& g\int\frac{d^3p}{(2\pi)^3}\frac{p^2}{3E} f^q\label{pressure} .
\end{eqnarray}
where $T$ and $\mu$ are the temperature and the chemical potential,
$V$ is the volume and  $g$ is the degeneracy factor.  
We have used the short-hand notation
\begin{equation}
\ln_q (x)\equiv \frac{x^{1-q}-1}{1-q} , \label{suba} 
\end{equation}
often referred to as q-logarithm.
It is straightforward to show that the relation
\begin{equation}
	\epsilon + P = Ts + \mu n
\end{equation}
(where $n, s$ refer to the densities of the corresponding quantities)
is satisfied.
Thermodynamics gives rise to the following  
differential relations:
\begin{eqnarray}
 d\epsilon = Tds + \mu dn,\label{a5}\\
dP = sdT + nd\mu.\label{a51}
\end{eqnarray}
\indent Since these are total differentials, thermodynamic consistency requires the following
Maxwell relations to be satisfied:
\begin{eqnarray}\label{a6}
 T &=& \left.\frac{\partial \epsilon}{\partial s}\right|_{n},\label{a61}\\
 \mu &=&\left.\frac{\partial \epsilon}{\partial n}\right|_{s},\\
 N &=& V\left.\frac{\partial P}{\partial \mu}\right|_{T},\label{a63}\\
 S &=& V\left.\frac{\partial P}{\partial T}\right|_{\mu}.\label{a64}
\end{eqnarray}
 It can be easily verified that this is indeed the case.\\
\indent  Following from Eq.~\eqref{number}, the momentum distribution is given by:
\begin{equation}
\frac{d^{3}N}{d^3p} = 
\frac{gV}{(2\pi)^3}
\left[1+(q-1)\frac{E -\mu}{T}\right]^{-q/(q-1)},
\label{tsallismu}
\end{equation}
or, expressed in terms of transverse momentum, $p_T$,  
transverse mass, $m_T \equiv \sqrt{p_T^2+ m ^2}$, and  rapidity  $y$  
\begin{equation}
\frac{d^{2}N}{dp_T~dy} = 
gV\frac{p_Tm_T\cosh y}{(2\pi)^2}
\left[1+(q-1)\frac{m_T\cosh y -\mu}{T}\right]^{-q/(q-1)} .
\label{tsallismu1}
\end{equation}
\indent At mid-rapidity, $y = 0$, and for zero chemical potential, as is relevant at 
the LHC, this reduces to: 
\begin{equation}
\left.\frac{d^{2}N}{dp_T~dy}\right|_{y=0} = 
gV\frac{p_Tm_T}{(2\pi)^2}
\left[1+(q-1)\frac{m_T}{T}\right]^{-q/(q-1)}.
\label{tsallisfit1}
\end{equation}
\indent In the limit where the parameter $q$ goes to 1 it is well-known that this reduces to 
the standard Boltzmann distribution~\cite{Tsallis:1987eu}.\\
\indent The parameterization given in Eq.~\eqref{tsallismu1} is close to
the one used by various collaborations~\cite{Abelev:2006cs,Adare:2010fe,Aamodt:2011zza,Aamodt:2011zj,Aad:2010ac,Khachatryan:2010us,Chatrchyan:2012qb}.
 The differences have been spelled out explicitly in~\cite{Cleymans:2013rfq} and will not be repeated here.
\section{Details of Transverse Momentum Distributions} 
%%%%%%%%%%%%
%
\indent The transverse momentum distributions of charged particles in $p-Pb$ collisions  were 
fitted using a sum of three Tsallis distributions, the 
first one for $\pi^+$, the
second one for $K^+$ and the third one for protons, $p$. The relative 
weights between these were 
 determined by the corresponding degeneracy factors, i.e. 1 for for $\pi^+$ and $K^+$
and 2 for  protons. 
The fit was taken at mid-rapidity and for $\mu = 0$ using the following expression:
\begin{equation}\label{tsallisfit}
   \left.  \frac{1}{2\pi p_{T}} \frac{d^{2}N_{\mathrm{charged \
particles}}}{dp_{T}dy}\right|_{y=0} = \frac{2V}{(2\pi)^{3}}
\sum\limits_{i=1}^{3} g_{i} m_{T,i}
\left[1+(q-1)\frac{m_{T,i}}{T}\right]^{-\frac{q}{q-1}},
\end{equation}
where $i=(\pi^{+},K^{+},p)$  and
$g_{\pi^{+}}=1$, $g_{K^{+}}=1$ and $g_{p}=2$. The factor $2$ in front
of the right hand side of this equation takes into account the
contributions of the antiparticles $(\pi^{-},K^{-},\bar{p})$.

\begin{figure}[ht]
\centering
\includegraphics[width=0.7\linewidth,height=8.0cm]{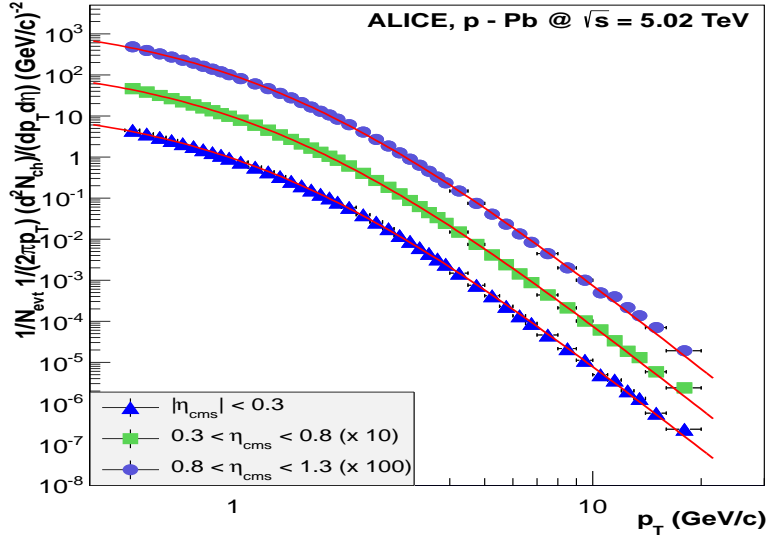}
\caption{Fits to transverse momentum distributions of charged particles in $p-Pb$
collisions at $\sqrt{s_{NN}}$ = 5.02 TeV measured by the ALICE collaboration~\cite{ALICE:2012mj} using the Tsallis distribution.}
\label{pt}
\end{figure}

\indent The Tsallis distribution  describes the transverse momentum distributions of charged particles 
in $p-Pb$ collisions as obtained by the ALICE collaboration~\cite{ALICE:2012mj} at
$\sqrt{s_{NN}}$ = 5.02 TeV in all pseudorapidity intervals remarkably well as shown in Fig.~\ref{pt}.\\
%%%%%%%%%%%%%%%%%%%%%%%%%%%%%%%%%%%%%%%%%%%%%%%%%%%%%%%%%%%
%
\indent The Tsallis parameter $q$  needed to describe the transverse momentum
distributions of charged particles is shown in Fig.~2. It is identical to the value obtained in $p-p$ 
collisions~\cite{Cleymans:2013rfq}.\\ 

The Tsallis parameter $T$  needed to describe the transverse momentum
distributions of charged particles is shown in Fig.~3. It is larger than the value obtained in $p-p$ 
collisions~\cite{Cleymans:2013rfq}. In order to clarify this it would be helpful to have data
at lower transverse momentum as the distribution is very sensitive to this region.

\indent The radius $R$ is given by: 
\begin{equation}
R \equiv {\left(\frac{3V}{4\pi}\right)}^{1/3} .
\end{equation}
\indent The values of $R$ are shown in Fig.~4 and are consistent (albeit slightly larger) than those deduced from 
$p-p$ collisions~\cite{Cleymans:2013rfq}.  No noticeable dependence on pseudorapidity
can be seen which could be due to the fact that the pseudorapidity region covered is always fairly central.

\begin{center}
\begin{table}
\begin{tabular}{|c|c|c|c|}\hline
Pseudorapidity Interval & q & T (MeV) & Radius (fm)\\ \hline
-0.3 $< \eta <$ 0.3  & 1.140 $\pm$ 0.001 & 112.81 $\pm$ 2.35  &  4.03 $\pm$ 0.12 \\
 0.3 $< \eta <$ 0.8  & 1.139 $\pm$ 0.001 & 113.35 $\pm$ 2.47  &  4.07 $\pm$ 0.13 \\
 0.8 $< \eta <$ 1.3  & 1.139 $\pm$ 0.001 & 111.92 $\pm$ 2.63  &  4.19 $\pm$ 0.14 \\
\hline
\end{tabular}
\caption{Values of the Tsallis parameters for $p-Pb$ collisions for various intervals of the pseudorapidity.}
\end{table}
\end{center}

\section{Conclusions}
\indent In conclusion, the Tsallis distribution,  Eq.~\eqref{tsallisfit1},  
leads to an  excellent description of the transverse momentum distributions
in high energy $p-Pb$ collisions as measured by the ALICE collaboration~\cite{ALICE:2012mj} at  $\sqrt{s_{NN}}$ = 5.02 TeV. 
Comparing the results to those obtained in $p-p$ collisions  it can be noted that the values of $q$ are compatible,
the values of $R$ are marginally larger while $T$ is substantially higher.
The values of $T$ and $R$ are very sensitive to the low $p_T$ part of the transverse momentum distribution
and extending the measurements to lower $p_T$ could bring much clarification here.\\
\indent The values obtained for the Tsallis parameter $q$, the temperature $T$  and the radius $R$ are consistent over all pseudorapidity intervals, a feature which does not 
become apparent when using 
other forms of parameterization.

\newpage
%%%%%%%%%%%%%%%%%%%%%%%%%%%%%%%%%%%%%%%%%%%%%%%%%%%%%%%%%%%
%with increasing  energy.
%
%
%
%
%%%%%%%%%%%%%%%%%%%%%%%%%%%%%%%%%%%%%%%%%%%%%%%%%%%%%%%%%%%
%%%%%%%%%%%%%%%%%%%%%%%%%%%%%%%%%%%%%%%%%%%%%%%%%%%%%%%%%%%
%
%
\begin{figure}
\centering
	\includegraphics[width=0.7\linewidth,height=8cm]{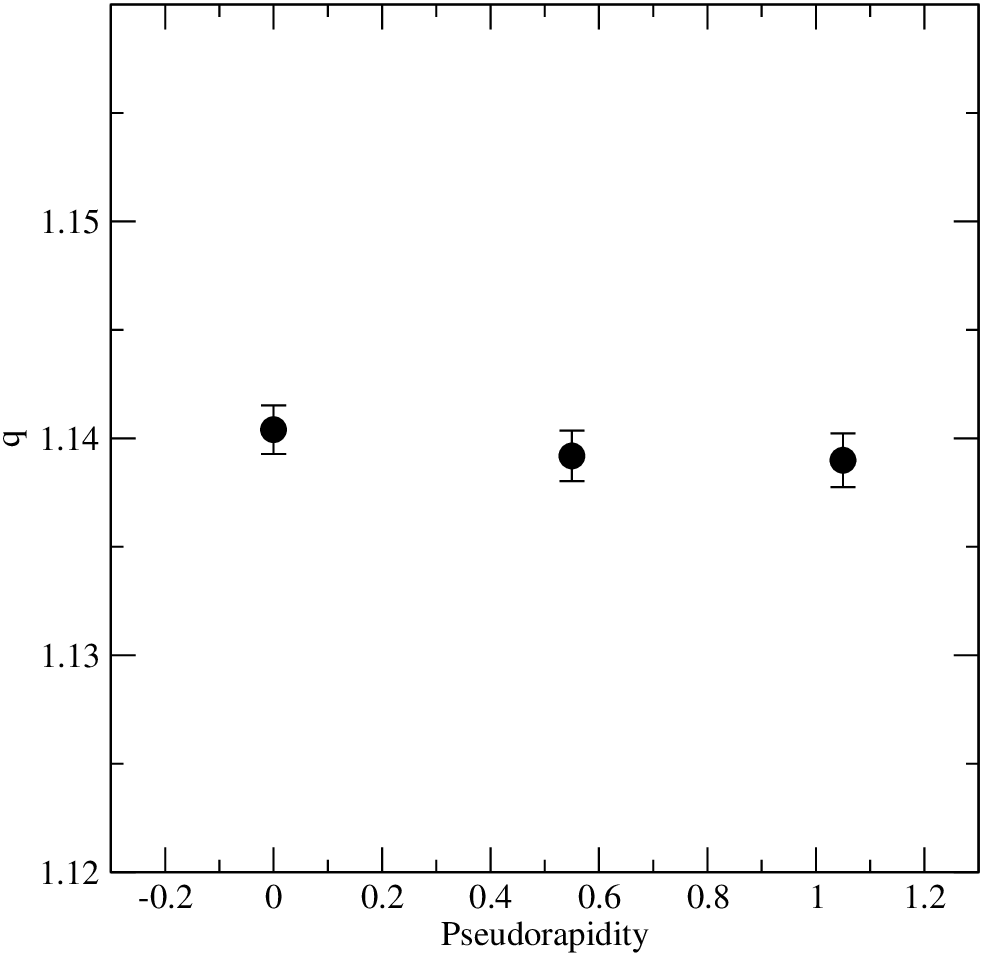}
\label{q_pPb}
\caption{Values of the Tsallis parameter $q$,
as a function of pseudorapidity interval, obtained from fits
to the transverse momentum distributions of charged particles obtained by the ALICE collaboration~\cite{ALICE:2012mj} in $p-Pb$ collisions
at $\sqrt{s_{NN}}$ = 5.02 TeV.}
\end{figure}
%
%
%
%%%%%%%%%%%%%%%%%%%%%%%%%%%%%%%%%%%%%%%%%%%%%%%%%%%%%%%%%%%
%
\begin{figure}[tbh]
\centering
	\includegraphics[width=0.7\linewidth,height=8cm]{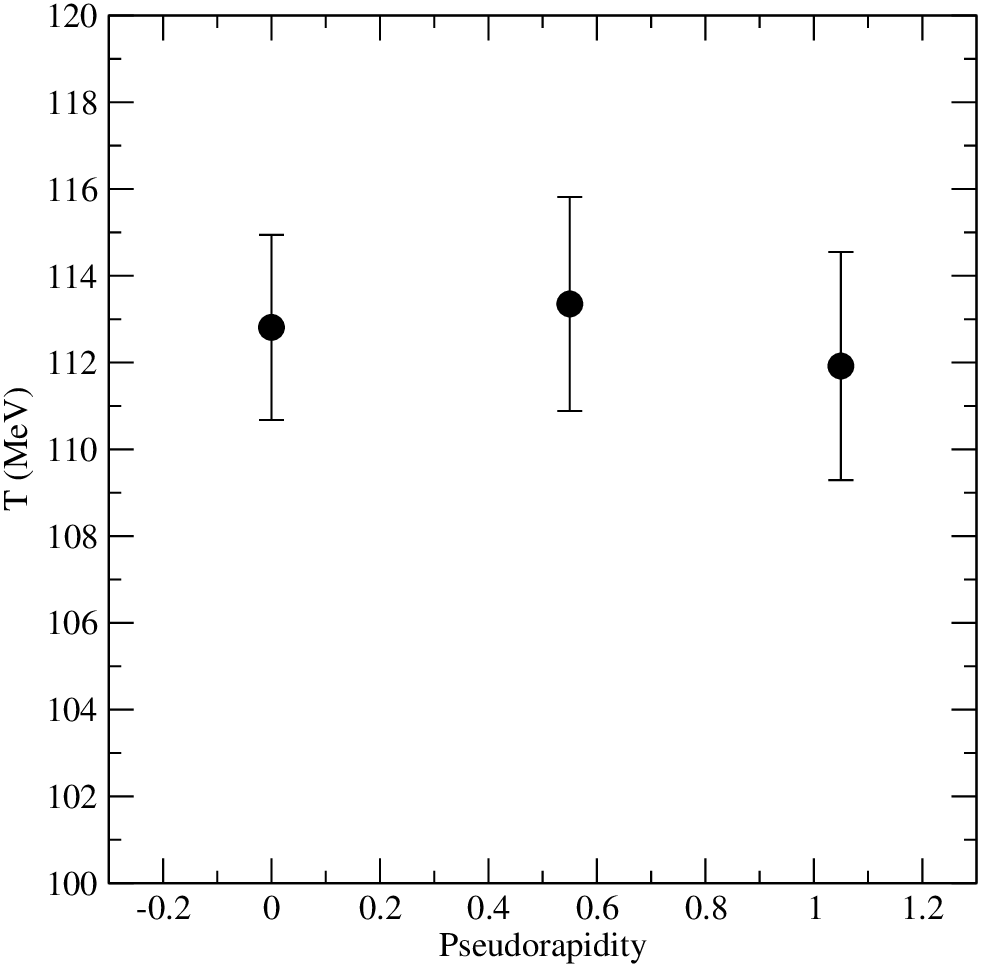}
\label{T_pPb}
\caption{Values of the Tsallis parameter, $T$,  
as a function of pseudorapidity interval, obtained from fits
to the transverse momentum distributions of charged particles  obtained by the ALICE collaboration~\cite{ALICE:2012mj} in $p-Pb$ collisions
at $\sqrt{s_{NN}}$ = 5.02 TeV.}
\end{figure}
%
%%%%%%%%%%%%%%%%%%%%%%%%%%%%%%%%%%%%%%%%%%%%%%%%%%%%%%%%%%%

%
\begin{figure}[tbh]
\centering
\includegraphics[width=0.7\linewidth,height=8cm]{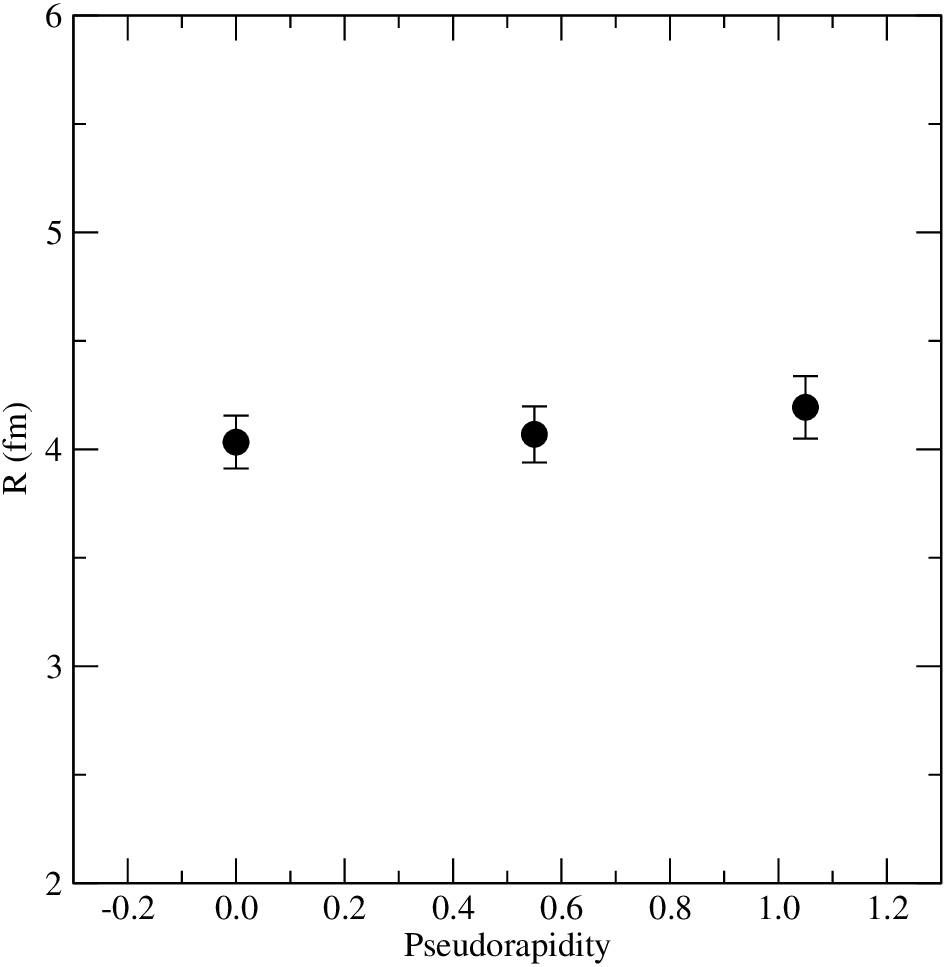}
\label{R_pPb}
\caption{Values of the Tsallis parameters $R$,
as a function of pseudorapidity interval, obtained from fits
to the transverse momentum distributions of charged particles  obtained by the ALICE collaboration~\cite{ALICE:2012mj} in $p-Pb$ collisions
at $\sqrt{s_{NN}}$ = 5.02 TeV.}
\end{figure}
%
%%%%%%%%%%%%%%%%%%%%%%%%%%%%%%%%%%%%%%%%%%%%%%%%%%%%%%%%%%%
%

\end{document}